\documentclass[12pt]{iopart}

\usepackage{amsfonts, amssymb, amsgen}
\usepackage{multicol}
\usepackage{braket}

\usepackage{cite}

\usepackage{bm}

\newcommand{\eq}[1]{\begin{equation} #1 \end{equation}}
\newcommand{\eqa}[2]{\begin{equation} #1 \label{#2} \end{equation}}
\newcommand{\balign}[1]{\begin{eqnarray} #1 \end{eqnarray}}

\newcommand{\todayd}{\the\year/\the\month/\the\day}

\newcommand{\bib}{\bibitem}

\newcommand{\lb}{\label}
\newcommand{\nt}{\nonumber}
\newcommand{\ft}[2]{\left. #1 \right|_{#2}}

\newcommand{\eqref}[1]{(\ref{#1})}

\newcommand{\bel}{\begin{easylist}}
\newcommand{\eel}{\end{easylist}}

\def \({\left(}
\def \){\right)}
\newcommand{\la}{\left\langle}
\newcommand{\ra}{\right\rangle}
\def \[{\left[}
\def \]{\right]}

\newcommand{\sumtwo}[2]%
{\mathop{\sum_{#1}}_{#2}}
\newcommand{\sumthree}[3]%
{\mathop{\mathop{\sum_{#1}}_{#2}}_{#3}}
\newcommand{\sumfour}[4]%
{\mathop{\mathop{\mathop{\sum_{#1}}_{#2}}_{#3}}_{#4}} 
\newcommand{\prodtwo}[2]%
{\mathop{\prod_{#1}}_{#2}}
\newcommand{\mintwo}[2]%
{\mathop{\min_{#1}}_{#2}}
\newcommand{\maxtwo}[2]%
{\mathop{\max_{#1}}_{#2}}
\newcommand{\maxthree}[3]%
{\mathop{\mathop{\max_{#1}}_{#2}}_{#3}}
\newcommand{\limtwo}[2]%
{\mathop{\lim_{#1}}_{#2}}
\newcommand{\suptwo}[2]%
{\mathop{\sup_{#1}}_{#2}}
\newcommand{\supthree}[3]%
{\mathop{\mathop{\sup_{#1}}_{#2}}_{#3}}
\newcommand{\supfour}[4]%
{\mathop{\mathop{\mathop{\sup_{#1}}_{#2}}_{#3}}_{#4}} 
\newcommand{\inftwo}[2]%
{\mathop{\inf_{#1}}_{#2}}
\newcommand{\infthree}[3]%
{\mathop{\mathop{\inf_{#1}}_{#2}}_{#3}}
\newcommand{\inffour}[4]%
{\mathop{\mathop{\mathop{\inf_{#1}}_{#2}}_{#3}}_{#4}} 

\newcommand\calC{{\cal C}}

\newcommand\calI{{\cal I}}
\newcommand\calJ{{\cal J}}

\newcommand{\bsa}{\boldsymbol{a}}

\newcommand{\bsp}{\boldsymbol{p}}

\newcommand{\bsx}{\boldsymbol{x}}

\newcommand{\bszero}{\boldsymbol{0}}

\newcommand{\Di}{\mathit{\Delta}}

\newcommand{\hsgm}{\hat{\sigma}}

\newcommand{\htau}{\hat{\tau}}
\newcommand{\dsgm}{\dot{\sigma}}

\newcommand{\hn}{\hat{n}}

\newcommand{\pss}{p^{\rm ss}}

\newcommand{\hcalJ}{\hat{\calJ}}
\newcommand{\hcalI}{\hat{\calI}}

\newcommand{\hcalC}{\hat{\calC}}

\def\rnum#1{\resizebox{0.5em}{\height}{\expandafter{\romannumeral #1}}}
\def\Rnum#1{\resizebox{0.5em}{\height}{\uppercase\expandafter{\romannumeral #1}}}

\begin{document}

\title[Fluctuation-response relation of time-symmetric quantities around general NESS]{Fluctuation-response relation of time-symmetric quantities around general nonequilibrium stationary state}

\author{Naoto Shiraishi}

\address{ Faculty of arts and sciences, University of Tokyo, 3-8-1 Komaba, Meguro-ku, Tokyo, Japan}
\ead{shiraishi@phys.c.u-tokyo.ac.jp}
\vspace{10pt}

\begin{abstract}
A connection between the response and fluctuation in general nonequilibrium stationary states is investigated.
We focus on time-symmetric quantities and find that the fluctuation of a kind of empirical measure can be expressed with the response of the empirical measure, current, and the time-symmetric current.
This relation is proven by using the fictitious stalling decomposition:
We decompose a single observed transition (edge in the state space) between two microscopic states into two transitions such that one of the transitions stalls in this stationary state.
Through this trick, relations for stalling stationary states apply to general nonequilibrium stationary states, which leads to the desired relation.

\end{abstract}

%
%
%
%
%

\section{Introduction}\label{intro}

The fluctuation-response relation around equilibrium states is one of the most important results in nonequilibrium statistical mechanics.
This relation connects two apparently independent quantities:
One is the conductivity against small external driving including heat conductivity and electrical conductivity.
The other is the equilibrium fluctuation of the current conjugate to the aforementioned external driving.
The fluctuation-response relation tells us that if we measure the response (conductivity) against external driving, then we immediately find the amount of the equilibrium fluctuation of the current without experiments, and vice versa.
In fact, the fluctuation-response relation was first found by Johnson in the experiment of the electric conduction and the fluctuation of electric current~\cite{Joh28}.
From the theoretical side, this relation was first explained by the consistency with the second law of thermodynamics~\cite{Nyq28}, and then formulated based on the microscopic foundation~\cite{Kub57}.

Bridging two different quantities is highly important in physics so that nontrivial conjectures are provided waiting for experimental verification.
The fluctuation-response relation clearly meets this desire, and to follow this success a number of studies have investigated fluctuation-response relations beyond the linear response regime.
One notable success is the discovery of the irreversible circulation~\cite{TT74}, which characterizes the violation of the Onsager reciprocity relation~\cite{Ons31} in nonequilibrium stationary states.
Another success is the Harada-Sasa relation~\cite{HS05}:
Although this relation applies only to Langevin systems, it clarifies the fact that the violation of the fluctuation-response relation is directly connected to the amount of stationary heat dissipation.
One traditional approach to the extension of the fluctuation-response relation, which is adopted by many papers, is the use of a time-evolution operator, with which we can obtain various formal extensions of the fluctuation-response relation~\cite{YK67, Aga72, Maes1, Maes2, PJP09, SS10, EMbook}.
However, in several cases, the obtained relations consist of complicated mathematical quantities whose physical meaning is not clear.

One pathway to avoid such formality is an approach with {\it inequalities} connecting the fluctuation and the response instead of equalities.
Since they are not equalities but bounds, there remain rooms to establish relations between established physical quantities.
One prominent example is the thermodynamic uncertainty relation~\cite{BS15, Ging16, GRH17} bounding stationary current fluctuations by entropy production.
The thermodynamic uncertainty relation is in fact a corollary of the fluctuation-response inequality~\cite{DS20}, which now becomes a standard technique to derive the thermodynamic uncertainty relation and its extensions~\cite{LGU20, KS20, Shi21, DS21}.
Besides them, many inequalities including response bounds for stationary distributions~\cite{OGH20} and trade-off relations between speed and dissipation~\cite{SST16, SFS18, SS19} can also be regarded as inequalities connecting fluctuation and response.

Most of the previous attempts to seek fluctuation-response relations treat currents or similar time-antisymmetric quantities (i.e., change its sign under time reversal).
Instead, in this paper, we shall focus on time-symmetric quantities (i.e., invariant under time reversal).
It is frequently claimed that time-symmetric quantities play key roles to understand nonequilibrium physics beyond the linear response regime~\cite{BM13, Mae20}.
We in particular treat the {\it time-symmetric current}~\cite{Shi22}, which is time-symmetric while its average takes the same value as that of the conventional (time-antisymmetric) current.
Very recently, it is shown that if the observed edge (a transition path between two microscopic states in interest) is stalling (i.e., the average probability current on this edge is zero), then the time-symmetric current satisfies the fluctuation-response relation even when the system is in a highly nonequilibrium stationary state~\cite{Shi22}.
This relation is violated as the system deviates from a stalling state.

In this paper, we extend the fluctuation-response relation for the time-symmetric current to general nonequilibrium stationary states beyond the stalling states.
We use two different expressions of the fluctuation-response relation:
One is with the time-symmetric current as mentioned above, and the other is with the {\it twisted empirical measure}, which consists of the empirical measures of two microscopic states.
The time-symmetric current and the twisted empirical measure have a simple relationship when the system stalls, while this connection no longer holds in general nonequilibrium stationary states.
We employ these two quantities and find that the fluctuation of the twisted empirical measure can be written in terms of the responses of the twisted empirical measure, the time-symmetric current, and the conventional current in general nonequiliburium stationary states.
This connection serves as an experimental probe of empirical measure fluctuation, which is uneasy to measure experimentally, through the measurement of responses of time-cumulative quantities.

We prove this relation by reducing it to the fluctuation-response relation in stalling states.
To this end, we decompose the observed edge into two edges $A$ and $B$ such that edge $A$ stalls.
We apply the fluctuation-response relation to edge $A$ and draw the desired relation from it.

\bigskip

This paper is organized as follows.
In \sref{s:setup}, we describe our setup and introduce key time-symmetric quantities: the time-symmetric current and the twisted empirical measure.
Using these time-symmetric quantities, we state our main result; the fluctuation-response relation around general nonequilibrium stationary states.
Since our proof of the main result heavily relies on the fluctuation-response relations in stall states, in \sref{s:stall} we explain these relations in detail.
These relations are derived through the Taylor expansion of the fluctuation theorems for two slightly-different partial entropy productions.
In \sref{s:gen}, we present the proof of our main result.
The key idea to prove this relation is the fictitious stalling method, which enables us to utilize results for stalling states in general nonequilibrium stationary states.

\section{Setup and main result}\lb{s:setup}

\subsection{Setup and key quantities}

Throughout this paper, we employ the framework of stochastic thermodynamics~\cite{Sei12, Shibook} and consider stationary continuous-time Markov jump processes on discrete states in $0\leq t\leq \tau$, where we finally take the long-time limit $\tau\to \infty$.
The time evolution of the probability distribution $\bsp$ follows the master equation
\eq{
\frac{d}{dt}\bsp(t)=R\bsp(t),
}
where $R$ is a transition matrix.
We suppose the local detailed-balance condition, that is, the entropy production rate $\dsgm(t)$ is expressed as
\eqa{
\dsgm(t)=\sum_{i,j}R_{ij}p_j(t)\ln \frac{R_{ij}p_j(t)}{R_{ji}p_i(t)}.
}{LDB}
Let $\la A \ra^{\tau}$ be an ensemble average of an observable $A$ in the stationary state in the time interval $0\leq t\leq \tau$.
We denote its long-time average by the bracket without superscript 
\eq{
\la A \ra:=\lim_{\tau \to \infty}\frac{1}{\tau}\langle A\rangle^\tau .
}

The cumulative probability current from state $i$ to another state $j$ is defined as 
\eq{
\hcalJ_{ij}:=\hn_{ij}-\hn_{ji},
}
where $\hn_{ij}$ is the number of jumps from state $j$ to $i$ and the hat symbol implies stochastic variables.
The current is a typical example of time-antisymmetric quantities in statistical mechanics.

We now proceed to time-symmetric quantities.
In Ref.~\cite{Shi22}, the cumulative {\it time-symmetric current} $\hcalI_{ij}$ defined as
\eqa{
\hcalI_{ij}:=R_{ij}\htau_j-R_{ji}\htau_i
}{def-I}
 is introduced, where 
 \eq{
 \htau_i:=\int_0^\tau dt a_{w(t), i}
 }
is the empirical staying time quantifying how long the state of the system $w(t)$ stays as state $i$.

We here notice two important properties of time-symmetric current.
First, the current and the time-symmetric current have the same ensemble average
\eq{
J_{ij}:=\langle \hcalJ_{ij}\rangle=\langle \hcalI_{ij}\rangle=:I_{ij}.
}
Second, by denoting the value of the cumulative current and the time-symmetric current with a trajectory $\Gamma$ by $\calJ_{ij}(\Gamma)$ and $\calI_{ij}(\Gamma)$ respectively, the time-reversal of the trajectory ($\Gamma\to \Gamma^\dagger$) changes the sign of the current, $\calJ_{ij}(\Gamma^\dagger)=-\calJ_{ij}(\Gamma)$, while it keeps the sign of the time-symmetric current, $\calI_{ij}(\Gamma^\dagger)=\calI_{ij}(\Gamma)$.

We denote the affinity from state $j$ to $i$ by
\eq{
x_{ij}:=\ln \frac{{R_{ij}}}{{R_{ji }}}.
}
If this transition is induced by a heat bath with inverse temperature $\beta$, we have $x_{ij}=\beta (E_j-E_i)$, where $E_i$ is the energy of state $i$.
If this transition is a chemical reaction induced by a particle bath with inverse temperature $\beta$ and chemical potential $\mu$, we have $x_{ij}=\beta \mu \Di n$, where $\Di n$ is the change in the number of particles through this reaction.
The time-symmetric current itself depends on $\bsx$ through $R_{ij}$ and $R_{ji}$ in $\hcalI$.
To emphasize this point, we also write $\hcalI_{ij}$ as $\hcalI_{ij, \bsx}$.

We further introduce another time-symmetric quantity named {\it twisted empirical measure}~\cite{SS15, Shi22} related to the time-symmetric current.
The twisted empirical measure is defined as
\eq{
\hcalC_{ij, \bsx}:=\frac{\htau_{j}}{\pss_{j}(\bsx)}-\frac{\htau_{i}}{\pss_{i}(\bsx)},
}
which quantifies the difference between empirical measures of states $i$ and $j$ relative to their averages.
Here, $\bsp^{\rm ss}(\bsx)$ is the stationary distribution with parameter $\bsx$.
If the edge $ij$ stalls (i.e., $R_{ji }(\bsx^*)\pss_{i}(\bsx^*)=R_{ij}(\bsx^*)\pss_{j}(\bsx^*)$ holds), the twisted empirical measure is connected to the time-symmetric current as
\eqa{
\hcalI_{ij,\bsx^*}=R_{ij}(\bsx^*)\pss_{j}(\bsx^*)\hcalC_{ij, \bsx^*}.
}{IC-connect}
On the other hand, if the edge $ij$ does not stall, the above relation no longer holds, and there is no simple relationship between the time symmetric current $\hcalI_{ij, \bsx}$ and the twisted empirical measure $\hcalC_{ij, \bsx}$.

\subsection{Main result}

For notational simplicity, we express the derivative of the current and the time-symmetric current on edge $ij$ around $\bsx=\bsx'$ as
\balign{
\langle \hcalJ\rangle'_{\bsx=\bsx'}&:=\ft{\frac{d \langle \hcalJ_{ij}\rangle_{\bsx}}{d x_{ij}}}{\bsx=\bsx'}, \\
\langle \hcalI_{ij, \bsx'}\rangle'_{\bsx=\bsx'}&:=\ft{\frac{d \langle \hcalI_{ij, \bsx'}\rangle_{\bsx}}{d x_{ij}}}{\bsx=\bsx'}.
}
In the latter relation, we need care in defining the derivative, since both the time-symmetric current itself and the path probability depend on affinity $x_{ij}$.
In this paper, we define the derivative such that the transition rate in the time-symmetric current is fixed at $\bsx'$, while that in the path probability is perturbed.
Similarly, we define the response of a twisted empirical measure around $\bsx=\bsx'$ as
\eq{
{\langle \hcalC_{ij, \bsx'}\rangle'}_{\bsx=\bsx'}:=\ft{\frac{d \langle \hcalC_{ij, \bsx'}\rangle_{\bsx}}{d x_{ij}}}{\bsx=\bsx'}.
}
We note that this derivative is equal to the derivative of the stochastic entropy difference between states $i$ and $j$:
\eqa{
{\langle \hcalC_{ij, \bsx'}\rangle'}_{\bsx=\bsx'}=\ft{\frac{d}{d {x_{ij}}} \ln \(\frac{\pss_{j}(\bsx)}{\pss_{i}(\bsx)}\)}{\bsx=\bsx'}.
}{dif-stoc-ent}

Now we shall connect these three responses to the fluctuation of the twisted empirical measure.
Around a general nonequilibrium stationary state with $\bsx= \bsx'$, we have the following equality
\eqa{
\frac{2}{\langle \hcalC_{ij, \bsx'}^2\rangle_{\bsx'}}=\frac{\langle \hcalI_{ij, \bsx'}\rangle'_{\bsx=\bsx'}-\langle \hcalJ_{ij}\rangle'_{\bsx=\bsx'} }{{\langle \hcalC_{ij, \bsx'}\rangle'}_{\bsx=\bsx'}},
}{main}
which is the main result of this paper.
As clearly seen, the fluctuation of empirical measures is directly connected to responses of time-cumulative quantities.
Remark that we can measure the right-hand side experimentally even with low time resolution, since these responses concern only time-cumulative quantities, which is tractable with low time resolution.
Recalling that measurement of the fluctuation of empirical measure requires high time resolution tracking, we find that this relation will serve as a tool to quantifying a kind of empirical measure fluctuation with accessible responses.

From the theoretical side, \eref{main} can be read as a relation that the ratio of the response and the fluctuation of the twisted empirical measure is equal to the discrepancy between the responses of the current and the time-symmetric current:
\eq{
\frac{{\langle \hcalC_{ij, \bsx'}\rangle'}_{\bsx=\bsx'}}{\langle \hcalC_{ij, \bsx'}^2\rangle_{\bsx'}}=\frac12 \( \langle \hcalI_{ij, \bsx'}\rangle'_{\bsx=\bsx'} - \langle \hcalJ_{ij}\rangle'_{\bsx=\bsx'}  \).
}

\section{Brief review on relations in stall states}\lb{s:stall}

Since our main result \eqref{main} is derived by the reduction to the result for stall states, in this section we describe fluctuation-response relations for stall states in detail.
Here, a stationary state is called stalling if the observed edge (the edge in interest) has zero average probability current (i.e., $\hcalJ_{ij}=0$).
We emphasize that a stall state may have a finite stationary probability current outside the observed edge.
This section serves as a quick review of the fluctuation-response relations for nonequilibrium stall states, which are shown in Refs.~\cite{APE16, PE17, Shi22}.

\subsection{Fluctuation-response relation on currents and time-symmetric currents}

Let $\bsx^*$ be a parameter with which the probability current between $i$ and $j$ stalls: $J_{ij}(\bsx^*)=0$.
In this setting, we have the following fluctuation-response relations on current $\hcalJ_{ij}$ and time-symmetric current $\hcalI_{ij,\bsx^*}$ around the stalling state with $\bsx=\bsx^*$:
\balign{
\ft{\frac{d \langle \hcalJ_{ij}\rangle_{\bsx}}{d x_{ij}}}{\bsx=\bsx^*}&=\frac12 \langle \hcalJ_{ij}^2\rangle_{\bsx^*}, \lb{J-main} \\
\ft{\frac{d \langle \hcalI_{ij, \bsx^*}\rangle_{\bsx}}{d x_{ij}}}{\bsx=\bsx^*}&=-\frac12 \langle \hcalI_{ij, \bsx^*}^2\rangle_{\bsx^*}, \lb{I-main}
}
The former relation \eqref{J-main} was derived by Altaner, Polletini, and Esposito~\cite{APE16}, and the latter relation \eqref{I-main} was derived by the author~\cite{Shi22}.
Remarkably, the same form of the fluctuation-response relations as the conventional fluctuation-response relation around equilibrium states are satisfied around nonequilibrium stationary states as long as the observed edge $ij$ stalls.

\subsection{Proof of Eqs.~\eqref{J-main} and \eqref{I-main}}

A standard derivation of the fluctuation-response relation around equilibrium states is the use of the Taylor expansion of the fluctuation theorem $\langle e^{-\hsgm}\rangle=1$~\cite{Nak11, Shibook}.
Similarly to this, two fluctuation-response relations, \eqref{J-main} and \eqref{I-main}, can be derived by the Taylor expansion of the fluctuation theorem for two different partial entropy productions; the original, or passive, partial entropy production~\cite{SS15} and the informed partial entropy production~\cite{PE17, Bis17}.
The partial entropy production is a generalization of entropy production to a subset of all possible transitions (edges in the state space).
The partial entropy productions have various applications from information thermodynamics~\cite{SIKS15, SMS16}, the efficiency of autonomous engines~\cite{Shi15, Shi17}, to the inference of dissipation~\cite{Bis17, Mar19}.
Both partial entropy productions satisfy fluctuation theorems, which lead to two different fluctuation-response relations on currents~\cite{SS15, APE16}.

Consider a system with parameter $\bsx$, where the edge $ij$ does not necessarily stall.
We first introduce two partial entropy productions.
The original partial entropy production with the edge $ij$ is defined as
\eqa{
\hsgm_{ij}:=a_{ij}(\bsx)\hcalJ_{ij}-J_{ij}(\bsx)\hcalC_{ij,\bsx},
}{def-pep}
where 
\eq{
a_{ij}(\bsx):=\ln \frac{{R_{ij}(\bsx)\pss_{j}(\bsx)}}{{R_{ji }(\bsx)\pss_{i}(\bsx)}}
}
is the total force associated with edge $ij$ with the stationary probability distribution $\pss$.
This partial entropy production satisfies the following fluctuation theorem~\cite{SS15}
\eqa{
\la e^{-\hsgm_{ij}}\ra_{\bsx}^\tau=1.
}{SS-FT}

We here remark on two facts on the partial entropy production.
First, $a_{ij}(\bsx)$ is a small parameter around the stalling state of the order of $\Di \bsx:=\bsx-\bsx^*$.
Second, the second term of the right-hand side of \eqref{def-pep} is evaluated as
\balign{
J_{ij}(\bsx)\hcalC_{ij, \bsx}&=(e^{a_{ij}(\bsx)}-1)R_{ji}(\bsx)\pss_i(\bsx)\hcalC_{ij, \bsx} \nt \\
&=a_{ij}(\bsx)\hcalI_{ij,\bsx}+O(a_{ij}(\bsx)^2),
}
which implies a useful expression of the partial entropy production:
\eq{
\hsgm_{ij}:=a_{ij}(\bsx)(\hcalJ_{ij}-\hcalI_{ij, \bsx})+O(a_{ij}(\bsx)^2).
}

Next, we introduce the informed partial entropy production with edge $ij$, which is defined as
\eq{
\hsgm_{ij}^{\rm I}:=\hcalJ_{ij}\ln \frac{R_{ij}(\bsx)\pss_{j}(\bsx^*)}{R_{ji }(\bsx)\pss_{i}(\bsx^*)}.
}
The informed partial entropy production appears very similar to the first term of the original partial entropy production \eqref{def-pep}, while the referred probability distribution is not $\bsx$ but $\bsx^*$.
Since the present system in interest is with parameter $\bsx$, not $\bsx^*$, to measure the informed partial entropy production experimentally we need to prepare another system where edge $ij$ stalls.
The informed partial entropy production also satisfies the fluctuation theorem~\cite{PE17}
\eqa{
\langle e^{-\hsgm_{ij}^{\rm I}}\rangle_{\bsx|\bsp^{\rm ss}(\bsx^*)}^\tau =1,
}{APE-FT}
where $\langle\cdot \rangle_{\bsx|\bsp(\bsx^*)}^\tau$ is the ensemble average with the transition rate $R(\bsx)$ starting from the initial distribution $\bsp(\bsx^*)$.

We now derive two fluctuation-response relations.
By expanding \eref{APE-FT} with $x_{ij}$  around the stalling state ($\bsx=\bsx^*$), the coefficients of $x_{ij}^2$ reads
\eqa{
\ft{\frac{d \langle \hcalJ _{ij}\rangle _{\bsx}}{d x_{ij}}}{\bsx=\bsx^*}=\frac12 \langle \hcalJ_{ij}^2\rangle_{\bsx^*},
}{APE1}
which is the fluctuation-response relation of currents \eqref{J-main}.
In addition, by expanding Eq.~\eqref{SS-FT} with $a_{ij}$  around the stalling state ($\bsx=\bsx^*$), the coefficients of $a_{ij}^2$ reads
\eq{
\ft{\frac{d}{d a_{ij}}(\langle \hcalJ_{ij}+ \hcalI_{ij, \bsx}\rangle_{\bsx})}{\bsa={\bszero}}=\frac12 (\langle (\hcalJ_{ij}+ \hcalI_{ij, \bsx^*})^2\rangle_{\bsx^*}). \lb{PEP1} 
}

\bigskip

In order to transform \eref{PEP1}, we use the following four relations.
\begin{enumerate}
\item For any $\bsx$, 
\eq{
\langle \hcalI_{ij, \bsx}\rangle_{\bsx}=0
}
is satisfied, which implies that the left-hand side of \eref{PEP1} equals $d \langle \hcalJ_{ij}\rangle/d a_{ij}|_{\bsa=\bszero}$.

\item The cross-correlation of the current and the time-symmetric current vanishes:
\eq{
\langle \hcalJ_{ij}\hcalI_{ij, \bsx^*}\rangle_{\bsx^*}=0.
}
In the case around equilibrium states, this relation is ensured by the fact that the observable $\hcalJ_{ij}\hcalI_{ij, \bsx^*}$ is time-antisymmetric.
In contrast, in the case around nonequilibrium stalling states, this relation is shown through straightforward but long calculation with the method of a counting field (see the Supplemental Material of \cite{Shi22}).

\item We transform the variable from $a_{ij}$ to $x_{ij}$.
The derivative is calculated as
\balign{
\ft{\frac{d a_{ij}}{d x_{ij}}}{\bsx=\bsx^*}&=\ft{\frac{d}{dx_{ij}}\[ x_{ij}+\ln \( \frac{\pss_l(\bsx)}{\pss_k(\bsx)}\)\]}{\bsx=\bsx^*} \nt \\
&=1+\frac{1}{R_{ij}(\bsx^*)\pss_l(\bsx^*)}\ft{\frac{d \langle \hcalI_{kl, \bsx^*}\rangle_{\bsx}}{d x_{ij}}}{\bsx=\bsx^*}.
}

\item The derivative of $\langle \hcalJ_{ij}\rangle_{\bsx}=R_{ij}(\bsx)\pss_j(\bsx)-R_{ji}(\bsx)\pss_i(\bsx)$ with respect to $a_{ij}$ at the stalling state $\bsa=\bszero$ is directly calculated as
\balign{
\ft{\frac{d}{d a_{ij}}(\langle \hcalJ_{ij}\rangle_{\bsx})}{\bsa={\bszero }}&=\ft{\frac{d}{d a_{ij}}(R_{ij}(\bsx)\pss_j(\bsx)a_{ij}+O(a_{ij}^2))}{\bsa={\bszero }} \nt \\
&=R_{ij}(\bsx^*)\pss_j(\bsx^*).
}
\end{enumerate}

Combining these four relations, we finally have
\balign{
&\ft{\frac{d}{d x_{ij}}(\langle \hcalJ_{ij}\rangle_{\bsx})}{\bsx=\bsx^*} \nt \\
=&\frac{d a_{ij}}{d x_{ij}}\ft{\frac{d}{d a_{ij}}(\langle \hcalJ_{ij}\rangle_{\bsx})}{\bsa={\bszero }} \nt \\*
=&\ft{\frac{d}{d a_{ij}}(\langle \hcalJ_{ij}\rangle_{\bsx})}{\bsa={\bszero }}+\ft{\frac{d}{d a_{ij}}(\langle \hcalJ_{ij}\rangle_{\bsx})}{\bsa={\bszero }}\frac{1}{R_{ij}(\bsx^*)\pss_l(\bsx^*)}\ft{\frac{d \langle \hcalI_{ij, \bsx^*}\rangle_{\bsx}}{d x_{ij}}}{\bsx=\bsx^*} \nt \\*
=&\ft{\frac{d}{d a_{ij}}(\langle \hcalJ_{ij}\rangle_{\bsx})}{\bsa={\bszero }}+R_{ij}(\bsx^*)\pss_j(\bsx^*)\frac{1}{R_{ij}(\bsx^*)\pss_l(\bsx^*)}\ft{\frac{d \langle \hcalI_{ij, \bsx^*}\rangle_{\bsx}}{d x_{ij}}}{\bsx=\bsx^*} \nt \\*
=&\frac12 (\langle \hcalJ_{ij}^2 \rangle_{\bsx^*}+ \langle \hcalI_{ij, \bsx^*}^2\rangle_{\bsx^*}) + \ft{\frac{d \langle \hcalI_{ij, \bsx^*}\rangle_{\bsx}}{d x_{ij}}}{\bsx=\bsx^*}.
}
Subtracting \eref{J-main} from the above relation, we obtain the desired result \eqref{I-main}.

\section{Proof of \eref{main}}\lb{s:gen}

\subsection{Main proof idea: fictitious stalling method}

Our key idea to prove \eref{main} is the {\it fictitious stalling method}, where we decompose a (non-stalling) edge into two edges such that one of the edges stalls.
Through this trick, we can apply results for stalling edges to general non-stalling edges in a nonequilibrium stationary state.

We decompose edge $ij$ in interest fictitiously into two edges $A$ and $B$.
The transition rates of these two edges are decomposed as
\balign{
R_{ji}({x_{ij}})=&R^A_{ji}({x_{ij}})+R^B_{ji}, \\
R_{ij}({x_{ij}})=&R^A_{ij}({x_{ij}})+R^B_{ij}.
}
Here, we set the transition rates in $B$ fixed independent of ${x_{ij}}$.
We further require that the edge $A$ stalls at $\bsx=\bsx'$:
\eq{
R^A_{ji}({x_{ij}})p_i({x_{ij}})=R^A_{ij}({x_{ij}})p_j({x_{ij}}).
}
Note that given a nonequilibrium stationary state, we have freedom of the choice of $R^A_{ji}({x'_{ij}})$ in the decomposition.
We call the description with two edges $A$ and $B$ as {\it fictitious description} and that with a single edge $ij$ as {\it original description}.

Applying results on a stalling edge to the edge $A$, we can derive \eref{main} in general nonequilibrium steady states.

\subsection{Proof}

Henceforth, we drop ${x_{ij}}$- and $\bsx$-dependence in case of no confusion.

We denote the affinity of edge $A$ by
\eq{
x_{ij}^A:= \ln \frac{R_{ij}^A}{R_{ji}^A}.
}
Suppose that we apply a small perturbation on edge $ij$ by $\Di x_{ij}$ (in the original description), which induces the change in the transition rates $\Di R_{ij}$ and $\Di R_{ji}$.
Since edge $B$ is fixed, all the changes in the transition rates are put on the change in the transition rates on edge $A$.
With this perturbation, the change in $x_{ij}^A$ is calculated as
\balign{
\Di x_{ij}^A&=\ln \frac{R_{ij}^A+\Di R_{ij}}{R_{ji}^A+\Di R_{ji}}-\ln \frac{R_{ij}^A}{R_{ji}^A} \nt \\
&=\frac{\Di R_{ij}}{R^A_{ij}} - \frac{\Di R_{ji}}{R^A_{ji}}+O(\Di x_{ij}^2) \nt \\
&=D\Di {x_{ij}} +O(\Di x_{ij}^2), \lb{DixijA}
}
where we defined
\eq{
D:=\frac{1}{R^A_{ij}(\bsx')}\ft{\frac{d}{d x_{ij}}R_{ij}(\bsx)}{\bsx=\bsx'}-\frac{1}{R^A_{ji}(\bsx')}\ft{\frac{d}{d x_{ij}}R_{ji}(\bsx)}{\bsx=\bsx'}
}
From \eref{DixijA}, we easily see
\eqa{
\frac{d x_{ij}^A}{d x_{ij}}=D.
}{xAx-dif}

We shall rewrite the fluctuation-response relation \eqref{I-main} around a stall state in terms of the twisted empirical measure.
Since the response and the fluctuation of the time-symmetric current are written as
\balign{
{\langle \hcalI_{ij, \bsx^*}\rangle'}_{\bsx=\bsx^*} &=R_{ij}(\bsx^*)\pss_j(\bsx^*){\langle \hcalC_{ij, \bsx^*}\rangle'}_{\bsx=\bsx^*}, \\
\langle \hcalI_{ij, \bsx^*}^2\rangle_{\bsx^*}&=(R_{ij}(\bsx^*)\pss_j(\bsx^*))^2 \langle \hcalC_{ij, \bsx^*}^2\rangle_{\bsx^*},
}
we find that \eref{I-main} is expressed as
\eqa{
R_{ij}(\bsx^*)\pss_{j}(\bsx^*)=-2\frac{{\langle \hcalC_{ij, \bsx^*}\rangle'}_{\bsx=\bsx^*}}{ \langle \hcalC_{ij, \bsx^*}^2\rangle_{\bsx^*}}.
}{main-stall}
Applying this relation to edge $A$, we have
\eqa{
R^A_{ij}(\bsx')\pss_{j}(\bsx')=-2\frac{1}{\langle \hcalC_{ij, \bsx'}^2\rangle_{\bsx'}}\ft{\frac{d \langle \hcalC_{ij, \bsx'}\rangle_{\bsx}}{d x^A_{ij}}}{\bsx=\bsx'}.
}{pf-mid1}
Using \eref{xAx-dif}, the derivative in the right-hand side is transformed into
\eqa{
\ft{\frac{d \langle \hcalC_{ij, \bsx'}\rangle_{\bsx}}{d x^A_{ij}}}{\bsx=\bsx'}=\ft{\frac{d x_{ij}}{d x_{ij}^A}\frac{d \langle \hcalC_{ij, \bsx'}\rangle_{\bsx}}{d x_{ij}}}{\bsx=\bsx'}=\frac{1}{D}{\langle \hcalC_{ij, \bsx'}\rangle'}_{\bsx=\bsx'}
}{pf-mid2}
Observe that
\balign{
R^A_{ij} \pss_j D &=R^A_{ij} \pss_j \( \frac{1}{R_{ij} }\frac{d R_{ij}}{d x_{ij}} - \frac{1}{R^A_{ji} }\frac{d R_{ji}}{d x_{ij}} \) \nt \\
&=\pss_j \frac{d R_{ij}}{d x_{ij}} - \pss_i \frac{d R_{ji}}{d x_{ij}} \nt \\
&=\( \frac{d R_{ij}\pss_j}{d x_{ij}} -  \frac{d R_{ji}\pss_i }{d x_{ij}} \) -\( R_{ij}\frac{d \pss_j }{d x_{ij}} - R_{ji}\frac{d \pss_i }{d x_{ij}}\) \nt \\
&=\langle \hcalJ_{ij}\rangle ' -\langle \hcalI_{ij, \bsx'}\rangle ' , \lb{pf-mid3}
}
where in the second line we used the stalling condition $R_{ij}^A\pss_j=R_{ji}^A\pss_i$.
Here, all the derivatives are taken at $\bsx=\bsx'$.
Plugging Eqs.~\eqref{pf-mid2} and \eqref{pf-mid3} into \eref{pf-mid1}, we arrive at the desired result \eqref{main}.

\bigskip

From this derivation, it is easy to see that our main result \eref{main} contains the fluctuation-response relation for the time-symmetric current around stalling states \eqref{I-main} as its corollary.
In fact, when edge $A$ stalls, we can set $R_{ij}^B=R_{ji}^B=0$ and $D=1$, which leads to
\eq{
\langle \hcalJ_{ij}\rangle ' -\langle \hcalI_{ij, \bsx'}\rangle ' =R_{ij}\pss_j .
}
Plugging this into \eref{main}, we arrive at \eqref{main-stall}, which is an alternative expression of the fluctuation-response relation for the time-symmetric current around stalling states \eqref{I-main}.

\section{Discussion}

We have shown that the fluctuation of the twisted empirical measure at any nonequilibrium stationary state is written in terms of the response of the twisted empirical measure, the time-symmetric current, and the conventional current.
This result elucidates a nontrivial role of time-symmetric quantities in the investigation of nonequilibrium stationary systems.
Previous studies state the importance of time-symmetric quantities in nonequilibrium stationary states~\cite{BM13, Mae20}, which is further supported by our finding.

The proof of our result relies on relations in nonequilibrium stalling states through a reduction technique.
Some relations in equilibrium states (e.g., fluctuation-response relation) also hold in nonequilibrium stalling states in the same form, while some others (e.g., Onsager reciprocity relation) need some modification~\cite{APE16, Shi22}.
This observation suggests that analyses on nonequilibrium stalling systems may serve as a good relay point in extending our understanding from equilibrium systems to general nonequilibrium stationary systems.

\ack
The author thanks Keiji Saito for the helpful discussion.
This work is supported by JSPS KAKENHI Grants-in-Aid for Early-Career Scientists Grant Number JP19K14615.

\end{document}